\documentclass[aps,prl,twocolumn,superscriptaddress]{revtex4}

\usepackage{graphicx}
\usepackage{amssymb}
\usepackage{times}
\usepackage{amsmath}
\usepackage{dcolumn}
\usepackage{bm}
\usepackage{colordvi}
\usepackage{color}
\usepackage{epsfig}

\begin{document}


\title{Realization of a monolithic high-reflectivity cavity mirror from a single silicon crystal}

\author{Frank Br{\"u}ckner}
\affiliation{Institut f{\"u}r Angewandte Physik, Friedrich-Schiller-Universit{\"a}t Jena, Max-Wien-Platz 1, 07743 Jena, Germany}

\author{Daniel Friedrich}
\affiliation{Albert-Einstein-Institut, Max-Planck-Institut f{\"u}r Gravitationsphysik and Leibniz Universit{\"a}t Hannover, Callinstr. 38, 30167 Hannover, Germany}

\author{Tina Clausnitzer}
\affiliation{Institut f{\"u}r Angewandte Physik, Friedrich-Schiller-Universit{\"a}t Jena, Max-Wien-Platz 1, 07743 Jena, Germany}

\author{Michael Britzger}
\affiliation{Albert-Einstein-Institut, Max-Planck-Institut f{\"u}r Gravitationsphysik and Leibniz Universit{\"a}t Hannover, Callinstr. 38, 30167 Hannover, Germany}

\author{Oliver Burmeister}
\affiliation{Albert-Einstein-Institut, Max-Planck-Institut f{\"u}r Gravitationsphysik and Leibniz Universit{\"a}t Hannover, Callinstr. 38, 30167 Hannover, Germany}

\author{Karsten Danzmann}
\affiliation{Albert-Einstein-Institut, Max-Planck-Institut f{\"u}r Gravitationsphysik and Leibniz Universit{\"a}t Hannover, Callinstr. 38, 30167 Hannover, Germany}

\author{Ernst-Bernhard Kley}
\affiliation{Institut f{\"u}r Angewandte Physik, Friedrich-Schiller-Universit{\"a}t Jena, Max-Wien-Platz 1, 07743 Jena, Germany}

\author{Andreas T{\"u}nnermann}
\affiliation{Institut f{\"u}r Angewandte Physik, Friedrich-Schiller-Universit{\"a}t Jena, Max-Wien-Platz 1, 07743 Jena, Germany}

\author{Roman Schnabel}
\affiliation{Albert-Einstein-Institut, Max-Planck-Institut f{\"u}r Gravitationsphysik and Leibniz Universit{\"a}t Hannover, Callinstr. 38, 30167 Hannover, Germany}

\date{\today}

\begin{abstract}
We report on the first experimental realization of a high-reflectivity cavity mirror that solely consists of a single silicon crystal. Since no material was added to the crystal, the urgent problem of 'coating thermal noise' that currently limits classical as well as quantum measurements is avoided. Our mirror is based on a surface nanostructure that creates a resonant surface waveguide. In full agreement with a rigorous model we realized a reflectivity of (99.79$\pm$0.01)\,\% at a wavelength of 1.55\,$\mu$m, and achieved a cavity finesse of 2784. We anticipate that our achievement will open the avenue to next generation high-precision experiments targeting fundamental questions of physics.
\end{abstract}


\maketitle

Cavity mirrors for laser radiation are essential as heavy test masses of space-time for the new field of gravitational wave astronomy \cite{Aufmuth,Smith}, as mechanical oscillators for targeting the quantum regime of macroscopic mechanical devices \cite{Kleckner,Schliesser,Mueller}, and for ultra-high-precision optical clocks designed for researching the nature of fundamental constants \cite{Udem,Young,Braxmaier,Numata}. Current limitations in all fields are set by the joint problem of lacking appropriate cavity mirror qualities. 
The purpose of cavity mirrors is to repeatedly retro-reflect laser light such that it constructively interferes with the stored cavity field yielding maximum field amplitudes and providing an output field of highest phase stability. In order to do so cavity mirrors need high reflectivities and a geometrically well-defined surface profile. If the mirror surface shows statistical fluctuations, for example driven by Brownian motion of the mirror's molecules, the phase fronts of subsequently reflected waves are slightly different and cannot perfectly interfere constructively, which results in a reduced cavity built-up and, most severely, in changes of the phase of the output laser beam. Motions of mirror surfaces, driven by thermal energy, are known as (\textit{Brownian}) \textit{thermal noise} and currently a major limiting factor in many research fields targeting fundamental questions of nature as mentioned above \cite{Levin,Braginsky,Gorodetsky}.

The best starting point for the fabrication of low thermal noise mirrors is to employ crystalline materials with high intrinsic mechanical quality factors (Q-factors), low thermal expansion coefficients and low absorption of the laser light, at cryogenic temperatures. A useful summary of thermal noise relations can be found in \cite{Gorodetsky}. Within the past years silicon was found to be a promising candidate \cite{Rowan} with an absorption of probably less than 10$^{-8}$\,cm$^{-1}$ at a 1550\,nm wavelength \cite{GKe95} and Q-factors of 10$^9$ at cryogenic temperatures  \cite{McGuigan}. In order to achieve high reflectivities for high-finesse setups dielectric multilayer coatings on the substrate's surface are currently employed, and reflectivities of up to 99.9998\,\% have been demonstrated \cite{Rempe}. However, recent theoretical and experimental research revealed that these coatings  reduce the substrate Q-factors and, most severely, lead to a strong inhomogeneous dissipation and therefore to a rapidly increasing Brownian thermal noise level \cite{Levin,Harry,Black,Nawrodt}. Thus, besides optimizing multilayer stacks \cite{Agresti} or trading off coherent thermal noise sources \cite{Kimble,Gorodetsky}, a coating-free (i.e. monolithic) mirror concept is of enormous interest. Previous published approaches such as corner reflectors \cite{Cella,Gossler} or whispering-gallery-mode resonators \cite{Matsko,Grudinin} are based on total internal reflection and significant optical path lengths inside a substrate giving rise to absorption and thermorefractive noise resulting from a temperature dependent index of refraction.


\begin{figure}[htb]
\centerline{\includegraphics[width=6.0cm]{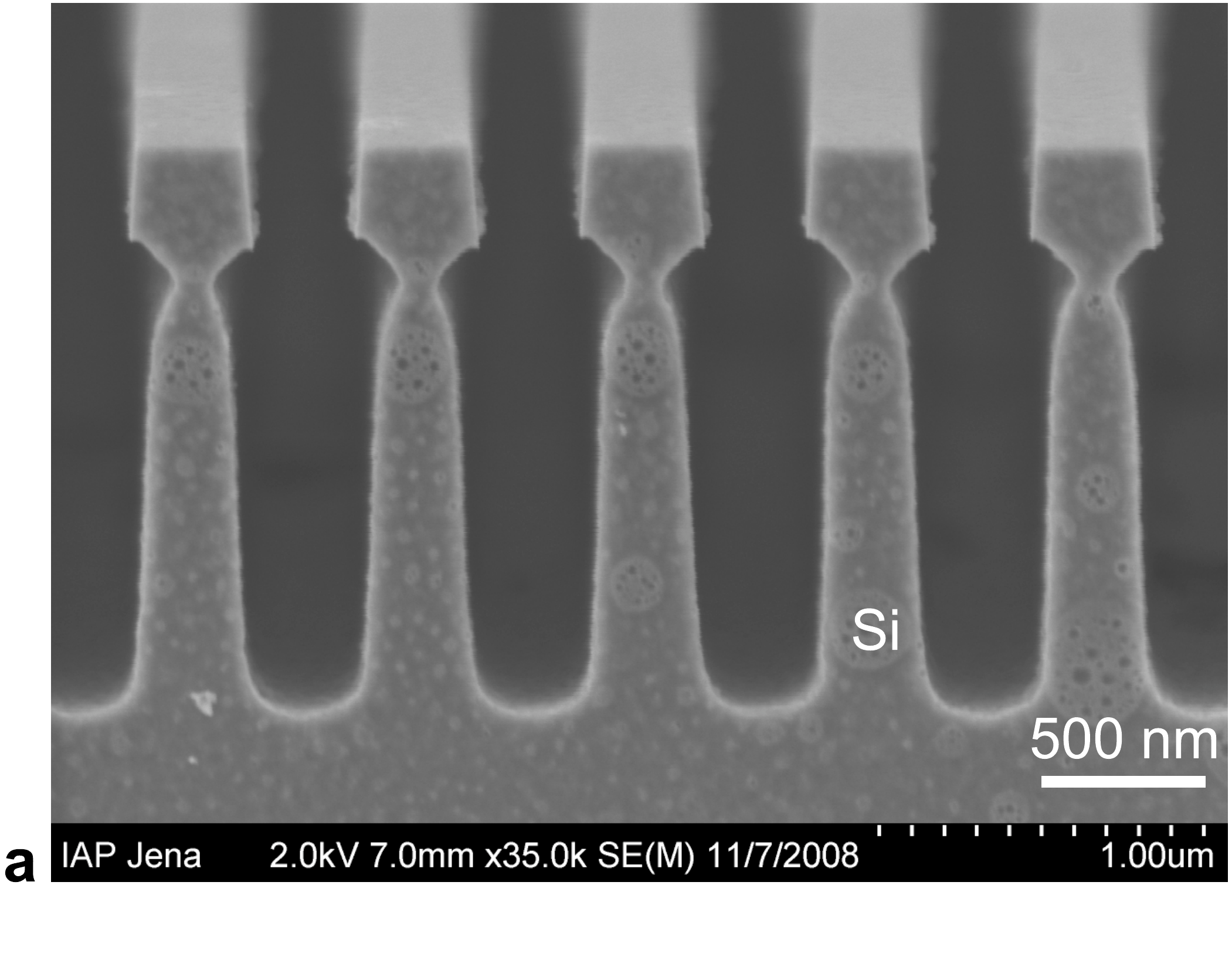}}
\centerline{\includegraphics[width=8.8cm]{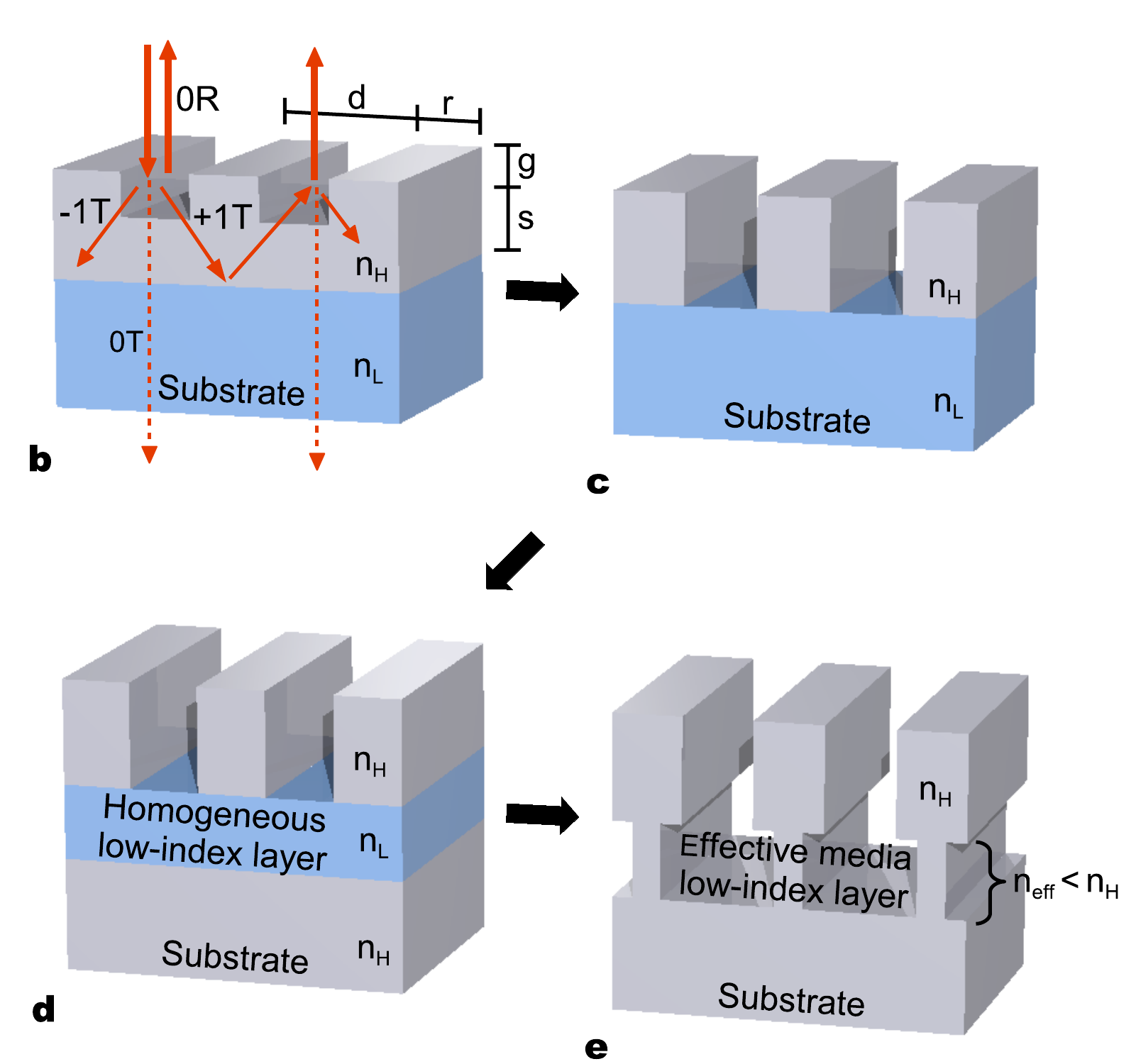}}
\caption{Monolithic mirror from a nanostructured single silicon crystal. (a) SEM (scanning electron microscope) cross-sectional view of a 700\,nm period T-shaped grating in a silicon bulk substrate that forms the monolithic cavity mirror's surface, efficiently reflecting normally incident light with a wavelength of 1.55\,$\mu$m. (b) Conventional resonant waveguide grating with a high-index layer corrugated at its surface on top of a low-index substrate. (c) Stand-alone high-index grating ridges corresponding to a zero waveguide layer thickness ($s =$ 0 nm). (d) Reduction of the low-index substrate to a thin layer. (e) For the monolithic implementation of the element in (d), the homogeneous low-index layer is replaced by an effective media low-index layer to advance the device to a monolithic reflector.}
\label{Fig1}
\end{figure}

This work experimentally demonstrates for the first time a monolithic surface mirror (see Fig.~\ref{Fig1}(a)), i.e.~a single piece of mono-crystalline silicon with a reflectivity high enough to form a laser cavity with a finesse of almost 3000. The achieved high reflectivity relies on resonant coupling to a guided optical mode of a surface nanostructure \cite{Golubenko,Sharon,Mateus}. Since no material is added to the silicon substrate, the currently limiting coating Brownian thermal noise as found in Ref. \cite{Levin,Harry} is avoided. A coating related reduction of substrate Q-factors should also be greatly avoided, as suggested by first experimental results \cite{Nawrodt}. Furthermore, our approach uses a \textit{broadband} guided optical mode and therefore does not increase the interaction length of light with matter thus keeping thermorefractive noise as well as absorption low. 

In Figs.~\ref{Fig1}(b)-(e) we plot the evolution from previous non-monolithic to the monolithic guided-mode resonant waveguide grating mirror. We use a simplified ray picture \cite{Sharon} in order to explain how reflectivities close to unity are achieved. The mirror architecture in Fig.~\ref{Fig1}(b) uses a second material coated on the substrate's surface, and was initially proposed for narrowband optical filters and switching applications in the mid 1980s. It comprises a periodically corrugated high-refractive index layer attached to a low-refractive index substrate. In order to allow for resonant reflection under normal incidence, the corrugation period must fulfill the following parameter inequalities, which can be derived from the well-known grating equation \cite{Bunkowski}:
\begin{eqnarray}
d &\!<\!& \lambda \quad \text{(to permit only zeroth order in air)} \,,\label{1} \\
\lambda/n_{\text{H}} &\!<\!& d  \quad \text{(first orders in high-index layer)} \,,\label{2}\\
d &\!<\!& \lambda/n_{\text{L}} \quad \text{(only zeroth order in substrate)} \,,\label{3}
\end{eqnarray}

\noindent where $d$ is the grating period, $\lambda$ is the light's vacuum wavelength and $n_{\text{H}}$ and $n_{\text{L}}$ are the higher and lower refractive indices, respectively. In our simplified ray picture, the first diffraction orders (-1T, +1T) within the high-index layer experience total internal reflection (at the interface to the low-index substrate) and, thus, can excite resonant waveguide modes propagating along the corrugated high-index layer. In turn, a certain fraction of the light inside the waveguide is coupled out again via the grating to both, the transmitted and reflected zeroth order (0T, 0R). If the grating period $d$, the groove depth $g$, the grating fill factor $f$ (ratio between ridge width $r$ and grating period $d$), and the high-index layer thickness $s$ with respect to the refractive index values of the involved materials are designed properly, all transmitted light can be prompted to interfere destructively, corresponding to a reflectivity of 100\,\%. Perfect reflectivity can also be achieved without the homogeneous part of the waveguide layer as proposed and realized in Ref. \cite{Bunkowski,Brueckner1}, see Fig.~\ref{Fig1}(c). The low-index substrate which is necessary for total internal reflection can be reduced to a layer \cite{Mateus}, see Fig.~\ref{Fig1}(d). This layer has to have a certain minimum thickness, for which evanescent transmission of the higher orders is still low. Although these approaches reduce the thick dielectric multilayer stack of conventional mirrors to a thin waveguide layer, at least one additional material has to be added still resulting in an increased mechanical loss.

Eventually, as shown in Fig.~\ref{Fig1}(e), we recently proposed to replace the remaining low-index layer by an \textit{effective} low-index layer \cite{Brueckner}. This grating layer exhibits the same period but has a lower fill factor (LFF) than the structure on top, and has an effective index $n_{\text{eff}} < n_{\text{H}}$. Since the high fill factor (HFF) grating on top does generate higher diffraction orders, referring to Ineq. (\ref{2}), the realization of the LFF grating as an effective medium without higher diffraction orders is not obvious \cite{Lalanne}. Only if the fill factor is sufficiently low, no higher diffraction orders are allowed to propagate as required, according to Ineq. (\ref{3}). The electric field inside this T-shaped structure can be expressed by discrete grating modes according to the so-called \textit{modal method} \cite{Botten}. If the fill factor is sufficiently low the LFF grating only supports the fundamental grating mode, which is related to the zeroth diffraction order in case of a homogeneous layer, whereas the HFF grating indeed allows for higher order modes to propagate. Similar to a conventional homogeneous layer, the remaining fundamental mode can show complete destructive interference for all light transmitted to the LFF grating. Thus, the monolithic T-shaped grating, as depicted in Fig.~\ref{Fig1}(e), can reach 100\,\% reflectivity for some wavelength if the fill factors and the groove depths of both gratings involved meet certain values. 

The design parameters of our T-shaped grating were found by rigorously solving the Maxwell equations \cite{Moharam}. We aimed for 100\,\% reflectivity for normal incidence of TM-polarized light (electric field vector perpendicular to the grating ridges) at 1550\,nm. The parameter set was further optimized to get a high first-order diffraction efficiency and, thus, broadband guided optical modes, as well as large parameter tolerances for the fabrication process. The result of our numerical calculations resulted in a grating period of 700\,nm and was presented in Ref. \cite{Brueckner}. We also rigorously calculated the optical near field distribution of this structure, see supplement \cite{EPAPS}.

\begin{figure}[htb]
\centerline{\includegraphics[width=7.5cm]{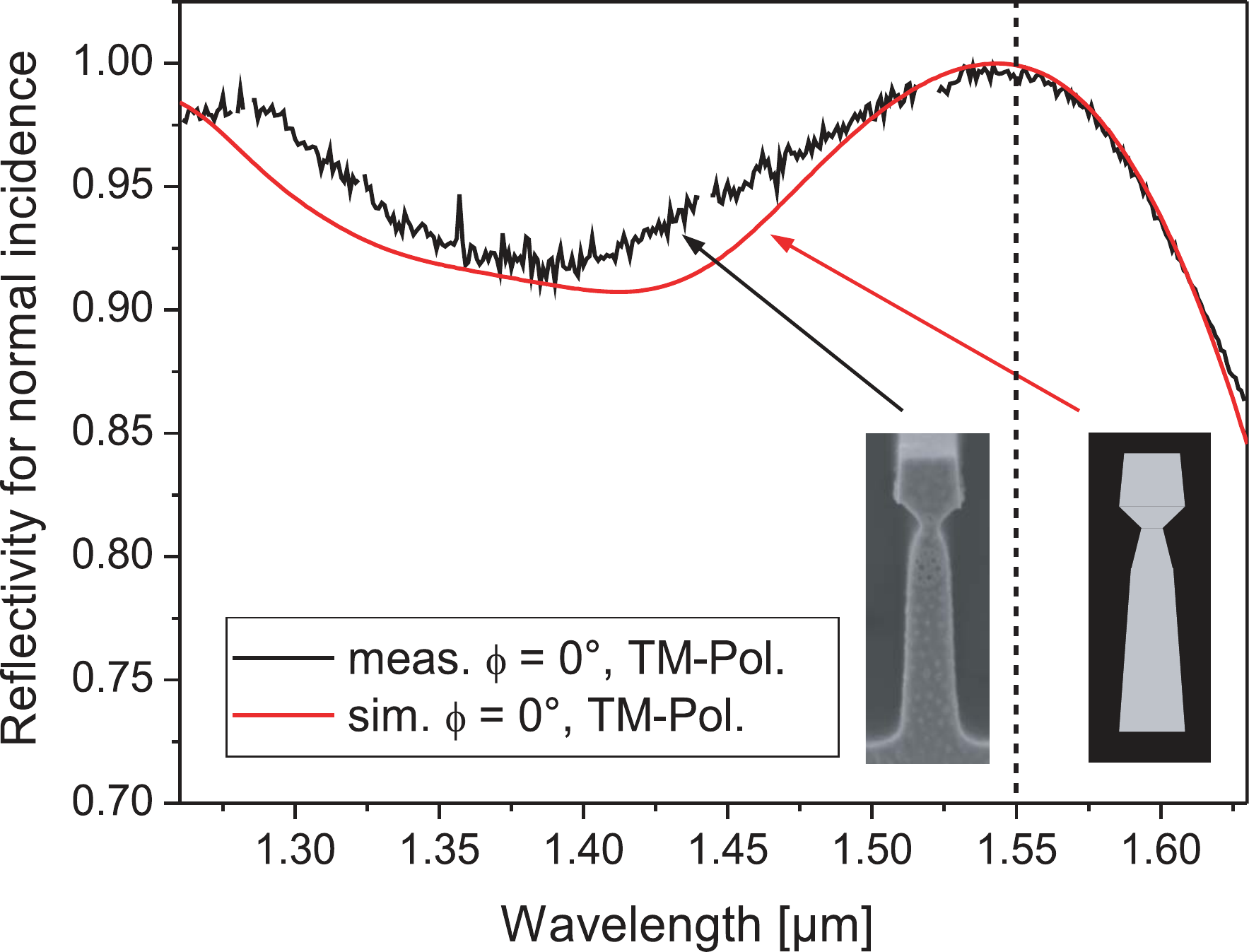}} \caption{Spectral reflectivity. Measured spectral reflectivity of the grating from Fig. \ref{Fig1}(a) for normal incidence $(\phi = 0\pm1)^\circ$ (black curve) and rigorously simulated spectral reflectivity for a grating profile approximating the real shape by a trapezoidal fragmentation (red curve).}
 \label{Fig2}
\end{figure}

For fabrication, a standard silicon wafer with 100\,mm in diameter was thermally oxidized with a 1\,$\mu$m silica layer and coated with a 80\,nm chromium layer, both serving as the mask during the silicon etching process. After spin-coating an electron beam sensitive (chemically amplified) resist on top, the 700\,nm period grating was defined by means of electron beam lithography for an area of (7.5 x 13)\,mm$^2$, aiming at a grating fill factor of 0.56 \cite{Brueckner}. The developed binary resist profile was then transferred into the chromium layer and subsequently into the oxidized silica layer as well as the silicon bulk substrate by utilizing an anisotropic (i.e. binary) ICP (Inductively-Coupled-Plasma) dry etching process. Here, the etching time was adjusted to match the desired groove depth of the upper silicon grating of about 350\,nm. The vertical grating groove side walls were then covered with a thin chromium layer by coating the whole device under an oblique angle. By using this technique, the groove side walls were protected from further ICP etching while the groove bottom remained accessible. A second, but this time isotropic (i.e. polydirectional), ICP etching process enabled the undercut of the upper grating to generate the low fill factor grating beneath. Here, a well-balanced ratio between horizontal and vertical etching rate played a decisive role to supply a sufficiently low grating fill factor ($<$\,0.3) as well as a minimum groove depth of the lower grating of about 500\,nm simultaneously \cite{Brueckner}. Finally, the etching mask materials (silica, chromium, and resist) were removed by means of wet chemical etching baring the mono-crystalline silicon surface structure.

Figure~\ref{Fig1}(a) depicts an SEM (scanning electron microscope) cross-sectional view on the fabricated mirror surface that has been characterized within this work. As expected, the shape of the grating ridges was not strictly rectangular, but it was within the parameter tolerances that predict high reflectivity \cite{Brueckner} (for another SEM image see supplement \cite{EPAPS}). 

\begin{figure}[htb]
\centerline{\includegraphics[width=8cm]{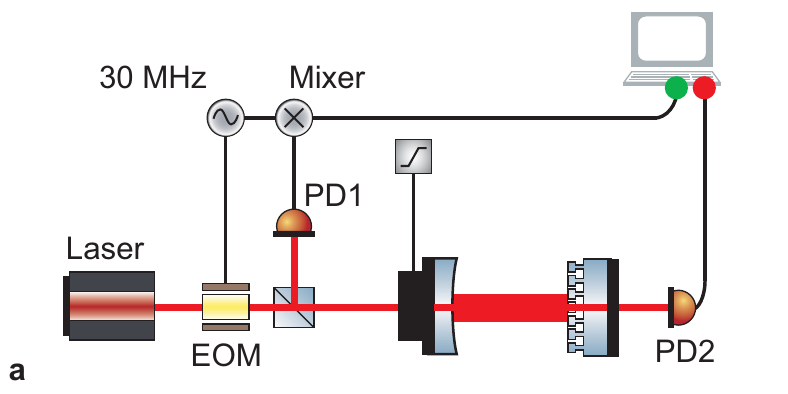}}
\centerline{\includegraphics[width=8cm]{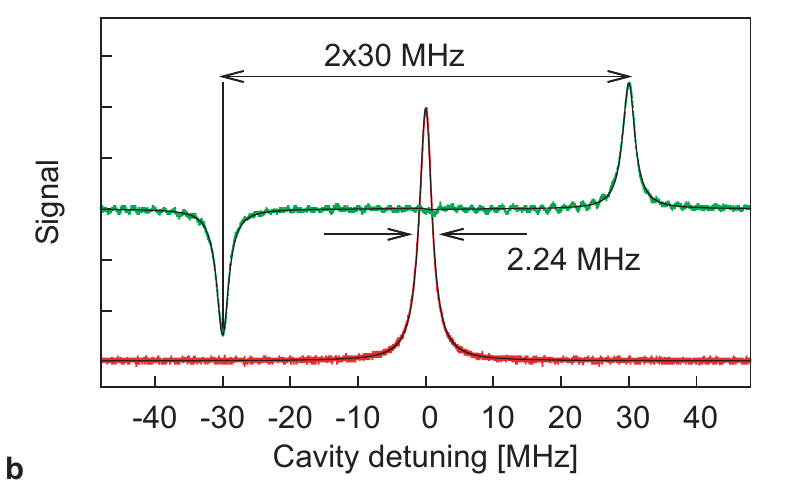}}
\caption{High-finesse cavity setup with monolithic end mirror. (a) Experimental setup for the characterization of the waveguide grating as a cavity end mirror. Electro-optical-modulator (EOM), photodiode (PD). (b) Scan over one cavity resonance (airy) peak (red line) with a linewidth of 2.24\,MHz measured in transmission (PD2), corresponding to a cavity finesse of 2784 and a power reflectivity of the monolithic mirror of 99.79\,\%. The cavity detuning was calibrated via the demodulated signal (frequency markers at $\pm30\,\mathrm{MHz}$, green line) in reflection of the cavity (PD1) that was generated with the PDH technique. Fitted theoretical lines are in black.}
 \label{Fig3}
\end{figure}

The first measurement of the mirror's reflectivity was performed under normal incidence $(0\pm1)^\circ$ and employed a fiber-coupled tunable diode laser. The measured data is shown in Fig.~\ref{Fig2} (black curve) and reveals a reflectivity of higher than 91.5\,\% for a rather broad spectral range from approximately 1.21\,$\mu$m to 1.61\,$\mu$m. The peak reflectivity is located close to the design wavelength of 1.55\,$\mu$m with a value of almost 100\,\%, where a measurement error of $\pm$\,0.3\,\% needs to be taken into account. The red curve in Fig.~\ref{Fig2} represents a rigorously simulated spectral response for a grating profile that has been formed by a trapezoidal fragmentation in order to approximate the real shape (indicated by the sketch on the bottom right hand side of Fig. \ref{Fig2}).

In order to demonstrate the high optical quality of our mirror we incorporated it as the end mirror in a standing-wave Fabry-Perot resonator, see Fig.~\ref{Fig3}(a). A conventional high quality multilayer coated mirror served as the coupling mirror with a measured power transmittivity of $\tau_1^2=(200\pm20)$\,ppm. Note that the monolithic mirror substrate had an unpolished rear surface and could not be used as the coupling mirror. By measuring the cavity's finesse $F$, this setup also enabled us to precisely determine the mirror's reflectivity under an angle of incidence of precisely zero degree at a wavelength of 1550\,nm. The product of the amplitude reflectivities $\rho_{12}=\rho_1 \rho_2$ of coupling mirror and end mirror, respectively, can be calculated from a measured finesse as follows
\begin{eqnarray}\label{4}
\rho_{12}=\rho_1 \rho_2 = 2-\cos{\frac{\pi}{F}}-\sqrt{\left(\cos{\frac{\pi}{F}}-2\right)^2-1}.
\end{eqnarray}

\noindent The finesse $F$ is defined as the ratio of free spectral range (FSR) $\nu_\mathrm{FSR}$ and cavity linewidth $\Delta \nu$ (full width at half maximum). The FSR was $\nu_\mathrm{FSR}=c/(2L)=(6.246\pm 0.13)\,\mathrm{GHz}$ with $c$ the speed of light and $L=(24\pm 0.5)$\,mm the distance between both mirrors. In order to obtain the linewidth of the linear Fabry-Perot resonator we used a calibrated tuning of the cavity length around an airy peak. The calibration was done via frequency markers at $\pm\,30\,\mathrm{MHz}$ around an airy peak using the Pound-Drever-Hall (PDH) technique \cite{Drever}, see Figs.~\ref{Fig3}(a),(b). For this purpose, a phase modulation was imprinted on the light by an electro-optical modulator (EOM). The detected signal in reflection (PD1) was then electronically demodulated by a local oscillator. We investigated 25 beam positions on the grating over an area of 4\,mm$^2$ with a beam size radius of $\approx\,50\,\mu$m. For each position, we did 12 measurements of the linewidth, which resulted in an averaged value of the power reflectivity of $\bar \rho_2^2=(99.7682\pm 0.0095)\%$ for the overall area. The smallest linewidth was determined to $\Delta\nu =$ (2.24 $\pm$ 0.07)\,MHz (see Fig.~\ref{Fig3}(b)). Hence, the finesse was found to be $F =$ 2784 $\pm$ 100, which corresponds to a waveguide grating power reflectivity of $\rho_2^2=(99.7945\pm 0.0086)\%$, referring to Eq. (\ref{4}) (for further information see supplement \cite{EPAPS}). Due to the unpolished back side of the substrate we could only set a lower limit on the transmission by means of power measurements of $\tau_2^2\geq (230\pm20)$\,ppm and, hence, an upper limit on optical losses due to absorption and scattering of $(1820\pm110)$\,ppm.


The optical reflectivity measured here is, to the best of our knowledge, the highest resonant reflection ever realized. The measured reflectivity of slightly below unity is attributed to deviations from the design parameters. However, it is in very good agreement with our rigorous model based on Maxwell equations that takes these deviations into account (Fig.~\ref{Fig2}).  Please note, that the same simulation predicts a reflectivity maximum for about 1.543\,$\mu$m. A verification in our cavity setup was not possible since the laser source was not tunable to this wavelength.\\
The reflectivity, as demonstrated here, in principle allows for a coating-free linear cavity with a finesse of $\approx$\,1500, which already reaches the regime of finesse values used in gravitational wave detectors. For example, the Advanced LIGO Fabry-Perot arm cavities are being designed for a finesse of a few hundred. Note that our technique can also be adapted to substrates with large radii of curvature, as required for long arm cavities. Our demonstrated monolithic mirror quality may also be already sufficient to provide an impact towards reaching the quantum regime of micro-mechanical oscillators. In Ref. \cite{Kleckner} optical cooling of a micro-mechanical oscillator down to 135\,mK was achieved with a finesse of only 200. Our demonstrated cavity linewidth of 2.24\,MHz is significantly smaller than many typical fundamental oscillator frequencies and the so-called good cavity regime can be reached \cite{Schliesser}.
Note that the cavity linewidth can be further reduced by increasing the cavity length. For applications in reference cavities and optical clocks \cite{Young,Braxmaier} the reflectivities should be further increased beyond the value demonstrated here, however, we expect that indeed considerable improvements towards a perfect reflectivity are possible with improved electron beam lithography and etching technologies. Having demonstrated the first monolithic surface mirror ever, our future work now aims for the realization of even higher reflectivities and for an in situ experimental confirmation that the thermal noise of our mirror concept is for fundamental reasons much lower than any other high-reflectivity mirror concept.


We acknowledge financial support from the Deutsche Forschungsgemeinschaft (DFG) within the Collaborativ Research centre TR7. We also acknowledge the centre of excellence QUEST for the allocation of the 1.55\,$\mu$m laser source. 


\begin{thebibliography}{99}

\bibitem{Aufmuth} P. Aufmuth and K. Danzmann, \emph{New J. Phys.} {\bf 7}, 202 (2005).

\bibitem{Smith} J. R. Smith {\it et al.} (for the LIGO Scientific Collaboration), \emph{Class. Quantum Grav.} {\bf 26}, 114013 (2009). 

\bibitem{Kleckner} D. Kleckner and D. Bouwmeester, \emph{Nature} {\bf 444}, 75-78 (2006).

\bibitem{Schliesser} A. Schliesser {\it et al.}, \emph{Nature Physics} {\bf 5}, 509-514 (2009). 

\bibitem{Mueller} H. M{\"u}ller-Ebhardt {\it et al.}, \emph{Phys. Rev. Lett.} {\bf 100}, 013601 (2008).

\bibitem{Udem} Th. Udem, R. Holzwarth, and T. W. H{\"a}nsch, \emph{Nature} {\bf 416}, 233-237 (2002).

\bibitem{Young} B. C. Young {\it et al.}, \emph{Phys. Rev. Lett.} {\bf 82}, 3799-3802 (1999).

\bibitem{Braxmaier} C. Braxmaier {\it et al.}, \emph{Phys. Rev. D} {\bf 64}, 042001 (2001).

\bibitem{Numata} K. Numata, A. Kemery, and J. Camp, \emph{Phys. Rev. Lett.} {\bf 93}, 250602 (2004).

\bibitem{Levin} Y. Levin, \emph{Phys. Rev. D} {\bf 57}, 659-663 (1998).

\bibitem{Braginsky} V. B. Braginsky, M. L. Gorodetsky, and S. P. Vyatchanin, \emph{Phys. Lett. A} {\bf 264}, 1-10 (1999).

\bibitem{Gorodetsky} M. L. Gorodetsky, \emph{Phys. Lett. A} {\bf 372}, 6813–-6822 (2008).

\bibitem{Rowan} S. Rowan {\it et al.}, \emph{Proc. SPIE} {\bf 4856}, 292-297 (2003). 

\bibitem{GKe95} M. A. Green and M. J. Keevers, \emph{Prog. Photovoltaic Res. Appl.} {\bf 3}, 189-192 (1995).

\bibitem{McGuigan} D. F. McGuigan {\it et al.}, \emph{J. Low Temp. Phys.} {\bf 30}, 621-629 (1978).

\bibitem{Rempe} G. Rempe {\it et al.}, \emph{Opt. Lett.} {\bf 17}, 363-365 (1992).

\bibitem{Harry} G. M. Harry {\it et al.}, \emph{Class. Quantum Grav.} {\bf 19}, 897-917 (2002).

\bibitem{Black} E. D. Black {\it et al.}, \emph{Phys. Lett. A} {\bf 328}, 1-5 (2004).

\bibitem{Nawrodt} R. Nawrodt {\it et al.}, \emph{New J. Phys.} {\bf 9}, 225 (2007).

\bibitem{Agresti} J. Agresti {\it et al.}, \emph{Proc. SPIE} {\bf 6286}, 628608 (2006).

\bibitem{Kimble} H. J. Kimble, B. L. Lev, and J. Ye, \emph{Phys. Rev. Lett.} {\bf 101}, 260602 (2008).

\bibitem{Cella} G. Cella and A. Giazotto, \emph{Phys. Rev. D} {\bf 74}, 042001 (2006).

\bibitem{Gossler} S. Go{\ss}ler {\it et al.}, \emph{Phys. Rev. A} {\bf 76}, 053810 (2007).

\bibitem{Matsko} A. B. Matsko {\it et al.}, \emph{J. Opt. Soc. Am. B} {\bf 24}, 1324-1335 (2007).

\bibitem{Grudinin} I. S. Grudinin, V. S. Ilchenko, and L. Maleki, \emph{Phys. Rev. A} {\bf 74}, 063806 (2006).

\bibitem{Golubenko} G. A. Golubenko {\it et al.}, \emph{Sov. J. Quantum Electron.} {\bf 15}, 886-887 (1985).

\bibitem{Sharon} A. Sharon, D. Rosenblatt, and A. A. Friesem, \emph{J. Opt. Soc. Am. A} {\bf 14}, 2985-2993 (1997).

\bibitem{Mateus} C. F. R. Mateus {\it et al.}, \emph{IEEE Phot. Techn. Lett.} {\bf 16}, 1676-1678 (2004).

\bibitem{Bunkowski} A. Bunkowski {\it et al.}, \emph{Class. Quantum Grav.} {\bf 23}, 7297-7303 (2006).

\bibitem{Brueckner1} F. Br{\"u}ckner {\it et al.}, \emph{Opt. Express} {\bf 17}, 163-169 (2009).

\bibitem{Brueckner} F. Br{\"u}ckner {\it et al.}, \emph{Opt. Lett.} {\bf 33}, 264-266 (2008).

\bibitem{Lalanne} P. Lalanne and D. Lemercier-Lalanne, \emph{J. Mod. Opt.} {\bf 43}, 2063-2085 (1996).

\bibitem{Botten} L. C. Botten {\it et al.}, \emph{Opt. Acta.} {\bf 28}, 413-428 (1981).

\bibitem{Moharam} M. G. Moharam and T. K. Gaylord, \emph{J. Opt. Soc. Am.} {\bf 71}, 811-818 (1981).

\bibitem{EPAPS} See EPAPS Document No. ... For more information on EPAPS, see http://www.aip.org/pubservs/epaps.html.

\bibitem{Drever} R. W. P. Drever {\it et al.}, \emph{Appl. Phys. B} {\bf 31}, 97-105 (1983).

\end{thebibliography}
\end{document}